\documentclass{article}
\newcommand{\apm}{\!\stackrel{\leftrightarrow\;}{\partial_{\mu}}\!}
\newcommand{\apn}{\!\stackrel{\leftrightarrow\;}{\partial'_{\nu}}\!}

\newcommand{\cint}{\int_{\Sigma}d\Sigma}

\title{Generalizations of normal ordering and  
applications to quantization in classical 
backgrounds}
\author{Hrvoje Nikoli\'c \\
Theoretical Physics Division, Rudjer Bo\v{s}kovi\'{c} Institute, \\
P.O.B. 180, HR-10002 Zagreb, Croatia \\
{\normalsize e-mail: hrvoje@thphys.irb.hr} \\
\makebox[1in]{} \\
}
\date{\today}
\begin{document}
\maketitle
%Suggested running head:
%Generalizations of normal ordering 

\begin{abstract}
%It is found that extraction of the positive (or the negative) frequency 
%part of a field at a point $x$ requires knowledge of the field on a 
%whole Cauchy surface at which $x$ lies. 
A nonlocal method of extracting 
the positive (or the negative) frequency part of a field, based 
on knowledge of a 2-point function, leads 
to certain natural generalizations of the normal ordering of quantum 
fields in classical gravitational and electromagnetic backgrounds 
and illuminates the origin of the recently discovered nonlocalities 
related to a local description of particles. 
A local description of
particle creation by gravitational backgrounds is given,
with emphasis on the case of black-hole evaporation.
The formalism  reveals 
a previously hidden relation between various definitions of the 
particle current and those of the energy-momentum tensor. 
The implications to particle creation by classical 
backgrounds, as well as  
to the relation between vacuum energy, dark matter, and cosmological
constant, are discussed.   
\end{abstract}
\vspace*{0.5cm}
%PACS numbers: 04.62.+v, 11.10.-z %\newline
KEY WORDS: Current of particle density ; classical background ;
particle creation
%\vspace*{0.5cm}

\section{Introduction}

The distinction between the positive frequency solutions and the negative 
frequency solutions of field equations plays a fundamental role in 
quantum field theory. In particular, their distinction is closely 
related to the distinction between annihilation and creation operators, 
which determine the representation of the field algebra, the particle 
content related to quantum fields, and the normal ordering related to the 
renormalization of various operators bilinear in fundamental fields.
However, the distinction between the positive and the negative frequencies 
has no invariant meaning in general relativity. In gauge theories, like 
electrodynamics, it also depends on the choice of gauge. This is closely 
related to the particle creation theoretically predicted to occur in 
classical gravitational \cite{bd,park,hawk} and electromagnetic 
\cite{man,padmprl} backgrounds. This is also closely related to 
the noncovariance (with respect to 
general coordinate transformations) of the concept of particles 
\cite{bd,ful,unruh}. The noncovariance is closely related to the fact 
that annihilation and creation operators are not local objects.
In particular, the horizon plays a fundamental role in the black-hole
evaporation \cite{hawk} and the Unruh effect \cite{unruh}, which 
raised serious doubts on the correctness of the formalism that 
describes these effects \cite{belin,belin2,belin3,nikolmpl}.  

Recently, progress in establishing the covariance of the concept 
of particles has been achieved by constructing an operator 
that represents the local current of 
particle density \cite{nikol_PLB,nikol_long}. This local current is 
covariant with respect to general coordinate transformations and invariant 
with respect to gauge transformations. Nevertheless, this local current 
possesses a nonlocal property related to the fact that the determination 
of the current at a point $x$ requires knowledge of the field on a
whole Cauchy surface at which $x$ lies. Besides, this current depends 
on the choice of a 2-point function. As shown in 
\cite{nikol_PLB,nikol_long}, there exist a choice that is consistent 
with the usual results of particle creation described by a Bogoliubov 
transformation \cite{bd,park,hawk,man,padmprl}.

In this paper we show that the nonlocality appearing in the calculation 
of the particle current is related to a nonlocal procedure of extracting 
the positive frequency part $\phi^+(x)$ and the 
negative frequency part $\phi^-(x)$ from the field $\phi(x)$. 
It appears  that the particle 
current can be written in a completely local form as an 
operator bilinear in the fields $\phi^+(x)$ and $\phi^-(x)$. By introducing 
certain generalizations of normal ordering, the current can be written 
even as an (suitably ordered) operator bilinear in the field $\phi(x)$. 
Different orderings of the particle current correspond to different 
orderings of other well-known local quantities, such as the energy-momentum 
tensor $T_{\mu\nu}(x)$. In particular, an ordering that retains the 
infinite vacuum energy  
leads also to an infinite number of particles in the vacuum. This suggests 
that it might not be meaningless to talk about the system in which the 
vacuum is at rest and, consequently, that this vacuum energy might not 
contribute to the cosmological constant. 

It also appears that the definitions of $\phi^+$ and $\phi^-$ depend 
on the choice of the 2-point function. Therefore, the normal ordering 
of $T_{\mu\nu}$ also depends on this choice. If the 2-point function 
is chosen such that the particle creation occurs, then $T_{\mu\nu}$ is 
not conserved. This suggests that one should choose the 
2-point function such that the particle creation does not occur.

In Sec.~\ref{NORMAL}, a method of extracting
$\phi^+$ and $\phi^-$ from a hermitian field $\phi$ is presented 
and used to define the normal ordering and some generalizations of it. 
The method is based on a particular choice of 2-point functions 
$W^+$ and $W^-$. This is applied in Sec.~\ref{PARTICLE} to 
write the particle current in a very elegant and purely local 
form. A generalization to other choices of $W^{\pm}$ is 
studied in Sec.~\ref{WW}, where the particle creation by a gravitational 
background is described in a local way, with emphasis on the description 
of particle creation by black holes. The formalism is generalized 
to complex scalar and spinor fields 
(interacting with a classical gravitational or 
electromagnetic background) in Sec.~\ref{COMPLEX}. 
The relation between the particle current and the corresponding 
energy-momentum tensor is discussed in Sec.~\ref{ENERGY}, where the 
implications to the particle creation 
and to the relation between vacuum energy, dark matter, and cosmological
constant, are discussed. The conclusions are drawn in Sec.~\ref{CONCL}. 

\section{Extraction of $\phi^{\pm}$ from $\phi$ and 
generalizations of normal ordering}\label{NORMAL}

A scalar hermitian field $\phi(x)$ in a curved background 
satisfies the equation of motion
\begin{equation}\label{8}
(\nabla^{\mu}\partial_{\mu}+m^2 +\xi R)\phi=0  .
\end{equation}
We choose a particular complete orthonormal set of solutions
$\{ f_k(x) \}$ of (\ref{8}) obeying the relations
\begin{eqnarray}\label{11}
& (f_k,f_{k'})=-(f^*_k,f^*_{k'})=\delta_{kk'} , & \nonumber \\
& (f^*_k,f_{k'})=(f_k,f^*_{k'})=0  , &
\end{eqnarray}
where the scalar product is defined as 
\begin{equation}\label{9}
(\phi_1,\phi_2)=i\cint^{\mu}\phi_1^* \apm \phi_2  .
\end{equation}
The field $\phi$ can be expanded as
\begin{equation}\label{10}
\phi(x)=\phi^+(x)+\phi^-(x) ,
\end{equation}
where
\begin{eqnarray}
\phi^+(x)=\sum_{k}a_k f_k(x) , & \nonumber \\ 
\phi^-(x)=\sum_{k}a^{\dagger}_k f^*_k(x).
\end{eqnarray}
Introducing the 2-point functions
\begin{eqnarray}\label{W+-}
W^+(x,x')=\sum_{k}f_k(x)f^*_k(x') , & \nonumber \\
W^-(x,x')=\sum_{k}f^*_k(x)f_k(x') ,
\end{eqnarray}
we find 
\begin{eqnarray}\label{phi+-}
\phi^+(x)=i\cint'^{\nu}W^+(x,x')\apn\phi(x') , & \nonumber \\
\phi^-(x)=-i\cint'^{\nu}W^-(x,x')\apn\phi(x') ,
\end{eqnarray}
which is the curved-spacetime generalization of the standard 
result for flat spacetime \cite{schweber}.
We see that the extraction of $\phi^+(x)$ and $\phi^-(x)$ from 
$\phi(x)$ is a nonlocal procedure. Note that the integrals in 
(\ref{phi+-}) do not depend on the choice of the timelike 
Cauchy hypersurface $\Sigma$ because $W^{\pm}(x,x')$ satisfy 
the equation of motion (\ref{8}) with respect to $x'$, just as 
$\phi(x')$ does. However, these integrals depend on the choice 
of $W^{\pm}(x,x')$, i.e., on the choice of the set $\{ f_k(x) \}$.

Having defined $\phi^+$ and $\phi^-$, one can define the normal 
ordering in the usual way as the ordering that puts $\phi^-$ on 
the left and $\phi^+$ on the right. Explicitly, 
\begin{equation}\label{normal:}
:\! \phi^+\phi^-\! :\, =\phi^-\phi^+,
\end{equation}
while the normal ordering of the combinations $\phi^-\phi^+$, 
$\phi^+\phi^+$, and $\phi^-\phi^-$ leaves these combinations 
unchanged. We generalize (\ref{normal:}) by introducing 4 
different orderings $N_{(\pm)}$ and $A_{(\pm)}$ defined by 
the relations analogous to (\ref{normal:}):
\begin{eqnarray}\label{orderings}
N_{(+)}\phi^+\phi^- =\phi^-\phi^+, \;\;\;\;
N_{(-)}\phi^+\phi^- =-\phi^-\phi^+,  & \nonumber \\
A_{(+)}\phi^-\phi^+ =\phi^+\phi^-, \;\;\;\;
A_{(-)}\phi^-\phi^+ =-\phi^+\phi^-.
\end{eqnarray}    
The normal ordering $N_{(+)}$ is identical to the normal ordering in
(\ref{normal:}). The normal ordering $N_{(-)}$ appears naturally 
in quantum field theory of fermion fields, but, as we shall see, 
it is also useful for boson fields. The antinormal orderings 
$A_{(\pm)}$ are useful because one can introduce the symmetric 
orderings $S_{(\pm)}$ defined by
\begin{eqnarray}\label{simorderings}
S_{(+)}=\frac{1}{2}[N_{(+)}+A_{(+)}] ,  & \nonumber \\
S_{(-)}=\frac{1}{2}[N_{(-)}+A_{(-)}].
\end{eqnarray}
When $S_{(+)}$ acts on a bilinear combination of fields,
then  it acts as the 
``default" ordering, i.e., $S_{(+)}\phi\phi=\phi\phi$. The 
usefulness of the $S_{(-)}$ ordering will become clear later.

\section{Particle current}\label{PARTICLE}   

The particle current for scalar hermitian fields can be written as 
\cite{nikol_PLB,nikol_long}
\begin{eqnarray}\label{16}
j_{\mu}(x) & = & \cint'^{\nu} \frac{1}{2} \{W^+(x,x')\apm \; \apn
\phi(x)\phi(x') \nonumber \\
 & & \;\;\;\;\;\;\;\;\;\;\;\;\;\,
+ W^-(x,x')\apm \; \apn \phi(x')\phi(x)\}.
\end{eqnarray}
Using (\ref{phi+-}), we see that it can be written in a 
purely local form as
\begin{equation}\label{loc1}
j_{\mu}(x)=\frac{i}{2}[\phi(x)\apm\phi^+(x)+\phi^-(x)\apm\phi(x)].
\end{equation}
Using (\ref{10}) and the identities 
$\phi^+\apm\phi^+ = \phi^-\apm\phi^- =0$, this can be written in a
very elegant form as
\begin{equation}\label{loc2}
j_{\mu}=i\phi^-\apm\phi^+ .
\end{equation}
Similarly, using (\ref{orderings}), this can be written 
in another elegant form without 
an explicit use of $\phi^+$ and $\phi^-$, as
\begin{equation}\label{loc3}
j_{\mu}=N_{(-)}\frac{i}{2}\phi\apm\phi .
\end{equation}
Note that the expression on the right-hand side of (\ref{loc3}) 
without the ordering $N_{(-)}$ vanishes identically. Nevertheless, 
the ordering $N_{(-)}$ makes this expression nonvanishing. 
This peculiar feature is probably the reason that the particle 
current has not been discovered earlier. 

The normal ordering $N_{(-)}$ provides that $j_{\mu}|0\rangle =0$. 
This is related to the fact that the total number of particles is
\begin{equation} 
N=\cint^{\mu}j_{\mu}=\sum_k a_k^{\dagger}a_k .
\end{equation}
Alternatively, one can choose the symmetric ordering $S_{(-)}$ 
defined in (\ref{simorderings}), i.e., one can define the 
particle current as
\begin{equation}\label{loc3S}
j_{\mu}=S_{(-)}\frac{i}{2}\phi\apm\phi .
\end{equation}
This leads to the total number of particles
\begin{equation}
N=\sum_k \frac{1}{2}(a_k^{\dagger}a_k+a_k a_k^{\dagger})=
\sum_k \left( a_k^{\dagger}a_k+\frac{1}{2} \right) .
\end{equation}
We see that this ordering generates the vacuum particle number equal to
$\sum_k 1/2$, in complete analogy with the vacuum energy which, in 
Minkowski spacetime, can be written as $\sum_k \omega_k/2$. We discuss 
the physical implications of this in Sec.~\ref{ENERGY}.

\section{Other choices of $W^{\pm}$ and particle creation}\label{WW}

When the gravitational background is time dependent, 
one can introduce a new set of solutions $u_l(x)$ for
each time $t$, such that $u_l(x)$ are positive-frequency modes
at that time. This leads to functions with an extra time dependence 
$u_l(x;t)$ that do not satisfy (\ref{8}) \cite{nikol_PLB,nikol_long}. 
Here $t$ is the time coordinate of the spacetime point
$x=(t,{\bf x})$.
We define $\phi^+$ and 
$\phi^-$ as in (\ref{phi+-}), but with the 2-point functions
\begin{eqnarray}\label{Wcreate}
W^+(x,x')=\sum_l u_l(x;t)u^*_l(x';t') , \nonumber \\
W^-(x,x')=\sum_l u^*_l(x;t)u_l(x';t') ,
\end{eqnarray}
used instead of (\ref{W+-}). As shown in \cite{nikol_PLB,nikol_long}, 
such a choice of the 2-point functions leads to a local description 
of particle creation consistent with the conventional global description 
based on the Bogoliubov transformation. Putting 
\begin{equation}
\phi(x)=\sum_k a_k f_k(x) + a^{\dagger}_k f^*_k(x)
\end{equation}
in (\ref{phi+-}) with (\ref{Wcreate}), we find  
\begin{equation}\label{phi+-t}
\phi^+(x)=\sum_l A_l(t) u_l(x;t), \;\;\;\; 
\phi^-(x)=\sum_l A^{\dagger}_l(t) u^*_l(x;t),
\end{equation}
where
\begin{equation}\label{e10}
A_l(t)=\sum_{k} \alpha^*_{lk}(t) a_k - \beta^*_{lk}(t) a^{\dagger}_k  ,
\end{equation}
\begin{equation}\label{e7}
\alpha_{lk}(t)=(f_k,u_l)  , \;\;\;\; \beta_{lk}(t)=-(f^*_k,u_l)  .
\end{equation}
By putting (\ref{phi+-t}) in (\ref{loc2}), we find
\begin{equation}\label{j_creation}
j_{\mu}(x)=i\sum_{l,l'}A^{\dagger}_l(t)  
u^*_l(x;t)\apm A_{l'}(t) u_{l'}(x;t).
\end{equation}
Note that, owing to the extra time dependence, the fields 
$\phi^+$ and $\phi^-$ in (\ref{phi+-t}) do not satisfy the 
equation of motion (\ref{8}). Consequently, the current 
(\ref{j_creation}) is not conserved, i.e., the quantity 
$\nabla^{\mu}j_{\mu}$ is a nonvanishing local scalar function 
describing the creation of particles in a local and invariant way, 
similarly as in \cite{nikol_PLB,nikol_long}.
%In general, since the extra time dependence influences
%the time derivative in (\ref{j_creation}),
%the number of particles at given $t$ differs from that obtained
%by conventional methods. As already discussed in
%\cite{nikol_PLB,nikol_long}, this difference corresponds to the fact
%that the conventional methods correspond to the adiabatic approximation,
%in which the number of particles does not change during a short
%period of time. Our formalism may be viewed as an improvement of
%this approximation. 

Let us now consider the questions {\em where} and {\em when} 
the particles are created. 
(Note that the space localization of the particle creation process 
cannot be directly considered in the conventional global 
approach based on the Bogoliubov transformation, simply because 
the local density of particles is not defined in this approach.) 
It is clear that $\nabla^{\mu}j_{\mu}(x)=0$ at the spacetime points 
$x$ at which the modes $u_l$ do not have the extra time dependence. 
Therefore, in general, the particles are created at the points at 
which the modes $u_l$ have this extra time dependence. 
One could choose the modes $u_l$ as highly nonlocal modes, such as 
the plane wave modes in Minkowski spacetime are. However, a question  
such as ``Where a particle with a definite momentum and a completely 
undetermined position is created?" does not make sense. Therefore, 
we assume that $u_l$ are some localized wave packets that, 
at a given instant of time, are negligible everywhere except in a small 
space volume \cite{aud}. 
Assume that $u_l(x)$ is a linear combination of modes 
that are all positive frequency modes at some instant of time. 
If the metric does not depend on time, then, during the time evolution,
these modes remain positive frequency modes. If the metric depends 
on time, then, during the time evolution, 
$u_l(x)$ ceases to be a linear combination of positive frequency modes. 
During an infinitesimal change of time, the modes $u_l(x)$ suffer 
an extra infinitesimal change related to the choice of new modes that are 
positive frequency modes at the new time. These infinitesimally 
modified new modes are also negligible everywhere except in the small 
(infinitesimally translated due to a finite group velocity of 
the packet) space volume. Therefore, the modes $u_l(x;t)$ have a 
nonnegligible extra time dependence only inside this small volume and only 
when the metric is time dependent. This implies that, in general, 
{\em the particles are created at the spacetime points at which the metric 
is time dependent}. 
%Note, however, that here ``time dependent"
%denotes dependence on a preferred global time coordinate $t$.
%Although in some cases there exists a natural choice of time,
%in general the ambiguity in choosing it is a problem both for our
%and for the conventional notion of particles. Some ideas on how
%this problem might be solved are discussed in
%\cite{nikol_long}. 
   
As a particular example, let us discuss the particle creation caused by 
a spherically symmetric gravitational collapse. Assume that all 
collapsing matter is contained in a ball with a radius $R(t)$. From the 
Birkhoff theorem it follows that the metric is 
time independent outside the ball, so all particles are created inside 
the ball. A certain amount of particles is created before the matter 
approaches a state in which all matter is trapped by an apparent 
horizon. These particles have not a thermal distribution. 
For the distribution to be approximately thermal, it is 
essential that the waves suffer an approximately exponential 
red shift, which occurs when the waves propagate close to the 
horizon. Therefore, since the particle production is a local 
process, the Hawking thermal radiation
results from particles that are created near the horizon.

The space components of the particle current $j_{\mu}$ determine 
the direction of the particle motion. 
Let us use them to confirm that the Hawking radiation is outgoing, as is 
usually argued by less direct arguments. Asymptotically, i.e., at 
late times and large distances from the horizon, we can  
approximate the modes $u_l$ with the usual plane wave modes. 
Therefore, in the asymptotic region we can make the replacement 
\begin{equation}\label{asimpu}
u_l(x;t)\rightarrow u_{\bf q}(x)=
\frac{e^{-iqx}}{\sqrt{V2\omega_{{\bf q}}}},
\end{equation}
where $\omega_{{\bf q}}=(m^2+{\bf q}^2)^{1/2}$.  
For convenience, $u_{\bf q}(x)$ are normalized in a finite volume $V$.
We integrate the current over the whole space. Since the integral is dominated 
by the contributions from the large distances, we use 
(\ref{j_creation}) and (\ref{asimpu}) to obtain
\begin{equation}\label{J}
J_{\mu}\equiv\int d^3 x\, j_{\mu}\simeq \sum_{{\bf q}} 
\frac{q_{\mu}}{\omega_{{\bf q}}} \, A_{{\bf q}}^{\dagger}A_{{\bf q}}.
\end{equation} 
In particular, in the vacuum $|0\rangle$ defined by $a_k|0\rangle =0$, 
(\ref{J}) and (\ref{e10}) give
\begin{equation}
\langle 0|J_{\mu}|0\rangle=\sum_{{\bf q}} \frac{q_{\mu}}{\omega_{{\bf q}}} \, 
n_{{\bf q}} ,
\end{equation}
where 
\begin{equation}\label{nq}
n_{{\bf q}}=\sum_k |\beta_{{\bf q}k}|^2 .
\end{equation}
The $\beta$-coefficients in (\ref{e10}) vanish for asymptotically ingoing 
modes because such modes have not experienced the black-hole gravitational 
field. Therefore, (\ref{nq}) has a form
\begin{equation}\label{nq2}
n_{{\bf q}}=\theta(q^r)\,n(q^r),
\end{equation}
where $q^r$ is the radial component of the 3-momentum ${\bf q}$ and 
$\theta$ is the step function. For massless fields, $n(q^r)$ is the 
thermal distribution equal to $[\exp(8\pi Mq^r)-1]^{-1}$, where 
$M$ is the mass of the black hole. We see that 
$\langle 0|J_0|0\rangle =\sum_{{\bf q}} n_{{\bf q}}$ 
represents the total number of produced particles. On the other hand, 
the Cartesian components $\langle 0|J_i|0\rangle$ vanish due to 
the cancellation of 
contributions from the opposite $q_i$'s. However, from (\ref{nq2}) we see 
that the radial component 
\begin{equation}
\langle 0|J^r|0\rangle =\sum_{{\bf q}} 
\frac{q^r}{\omega_{{\bf q}}}\, n_{{\bf q}}
\end{equation}
is positive, which confirms that the flux of created particles is 
outgoing.

%The choice of the 2-point functions (\ref{Wcreate}) depends on the choice 
%of the time coordinate. Therefore, in general, a natural choice 
%of the 2-point functions (\ref{Wcreate}) does not exist. In 
%\cite{nikol_PLB,nikol_long} an alternative choice has been introduced. 
%This is
%\begin{equation}\label{W=G}
%W^{\pm}(x,x')=G^{\pm}(x,x'),
%\end{equation}
%where $G^{\pm}(x,x')$ are determined by
%(but in general not equal to) the Schwinger-DeWitt Green
%function \cite{schw,dewitt}. Since the reasons for this choice
%are explained in detail in \cite{nikol_PLB,nikol_long}, here we
%only mention few important results derived there. In
%(\ref{W=G}), $G^{\pm}(x,x')$ is a modification of the
%Schwinger-DeWitt Green function valid
%even when $x$ is not close to $x'$. For $x=x'$ it can be written
%in the form of (\ref{W+-}), which reveals the positivity
%for $x=x'$. The functions
%$G^{\pm}(x,x')$ satisfy the equation of motion (\ref{8}).
%Consequently,
%the particle current in which $\phi^{\pm}$ are calculated by putting
%(\ref{W=G}) in (\ref{phi+-}) is conserved. This suggests that classical
%gravitational backgrounds might not create particles, which
%we discuss in more detail in Sec.~\ref{ENERGY}. Another consequence
%is the fact that the notion of particle defined in this way does
%not depend on the choice of a global time.

\section{Generalization to complex fields}\label{COMPLEX}

A complex scalar field $\phi(x)$ and its hermitian conjugate field
$\phi^{\dagger}(x)$ in an arbitrary gravitational background 
can be expanded as
\begin{equation}
\phi=\phi^{(P)+}+\phi^{(A)-}, \;\;\;\;
\phi^{\dagger}=\phi^{(P)-}+\phi^{(A)+},
\end{equation}
where
\begin{eqnarray}
\phi^{(P)+}(x)=\displaystyle\sum_k a_k f_k(x) , \;\;\;\;
\phi^{(P)-}(x)=\displaystyle\sum_k a^{\dagger}_k f^*_k(x) , \nonumber \\
\phi^{(A)+}(x)=\displaystyle\sum_k b_k f_k(x) , \;\;\;\;
\phi^{(A)-}(x)=\displaystyle\sum_k b^{\dagger}_k f^*_k(x) .
\end{eqnarray}
In a similar way as in Sec.~\ref{PARTICLE}, we find
\begin{eqnarray}
\phi^{(P)+}(x)=i\cint'^{\nu}W^+(x,x')\apn\phi(x') , & \nonumber \\
\phi^{(A)+}(x)=i\cint'^{\nu}W^+(x,x')\apn\phi^{\dagger}(x') , 
  & \nonumber \\
\phi^{(P)-}(x)=-i\cint'^{\nu}W^-(x,x')\apn\phi^{\dagger}(x') , 
  & \nonumber \\
\phi^{(A)-}(x)=-i\cint'^{\nu}W^-(x,x')\apn\phi(x') .
\end{eqnarray}
The particle current $j_{\mu}^{(P)}$ and the antiparticle current 
$j_{\mu}^{(A)}$ are \cite{nikol_PLB,nikol_long}
\begin{eqnarray}\label{curPA}
j^{(P)}_{\mu}(x) & = & \cint'^{\nu} \frac{1}{2} \{W^+(x,x')\apm \; \apn
\phi^{\dagger}(x)\phi(x') \nonumber \\
 & & \;\;\;\;\;\;\;\;\;\;\;\;\;\,
+W^-(x,x')\apm \; \apn \phi^{\dagger}(x')\phi(x) \} ,
\nonumber \\
j^{(A)}_{\mu}(x) & = & \cint'^{\nu} \frac{1}{2} \{W^+(x,x')\apm \; \apn
\phi(x)\phi^{\dagger}(x') \nonumber \\
 & & \;\;\;\;\;\;\;\;\;\;\;\;\;\,
+W^-(x,x')\apm \; \apn \phi(x')\phi^{\dagger}(x) \}.
\end{eqnarray}
Therefore, they can be written in a purely local form 
similar to (\ref{loc2}) as 
\begin{eqnarray}
j_{\mu}^{(P)}=i\phi^{(P)-} \apm \phi^{(P)+} + j_{\mu}^{{\rm mix}}, 
\nonumber \\
j_{\mu}^{(A)}=i\phi^{(A)-} \apm \phi^{(A)+} - j_{\mu}^{{\rm mix}},
\end{eqnarray}
where
\begin{equation}
j_{\mu}^{{\rm mix}}=\frac{i}{2}[\phi^{(P)-}\apm\phi^{(A)-} -
\phi^{(P)+}\apm\phi^{(A)+}].
\end{equation}

The current of charge $j_{\mu}^{(-)}$, defined as
\begin{equation}
j_{\mu}^{(-)}=j_{\mu}^{(P)}-j_{\mu}^{(A)},
\end{equation}
can be written in more familiar forms as \cite{nikol_PLB,nikol_long}
\begin{eqnarray}
j_{\mu}^{(-)} & = & \, :\! i\phi^{\dagger} \apm \phi \! : \nonumber \\
 & = & \frac{i}{2}[\phi^{\dagger} \apm \phi - \phi \apm \phi^{\dagger}].
\end{eqnarray}
Using (\ref{orderings}), we see that this can also be written as
\begin{eqnarray}\label{j-} 
j_{\mu}^{(-)} & = & N_{(+)} i\phi^{\dagger} \apm \phi \nonumber \\
 & = & N_{(+)} \frac{i}{2} [\phi^{\dagger} \apm \phi -
\phi \apm \phi^{\dagger} ] .  
\end{eqnarray}

The current of total number of particles $j_{\mu}^{(+)}$ is defined as 
\begin{equation}
j_{\mu}^{(+)}=j_{\mu}^{(P)}+j_{\mu}^{(A)}.
\end{equation}
It is shown in \cite{nikol_long} that $j_{\mu}^{(+)}$
can be written as a sum of two particle currents attributed to the hermitian 
fields $\phi_1$ and $\phi_2$ defined by 
\begin{equation}\label{phiherm}
\phi=\frac{\phi_1+i\phi_2}{\sqrt{2}},
\end{equation} 
as
\begin{equation}\label{j+}
j_{\mu}^{(+)}=j_{\mu}^{(1)}+j_{\mu}^{(2)},
\end{equation}
where $j_{\mu}^{(1)}$ and $j_{\mu}^{(2)}$ are two currents of the form
(\ref{16}). Therefore, using (\ref{loc3}), we can write (\ref{j+}) as 
\begin{equation}\label{j+2}
j_{\mu}^{(+)}=N_{(-)} \frac{i}{2} [\phi_1\apm\phi_1
+\phi_2\apm\phi_2 ] .
\end{equation}
Using (\ref{phiherm}), it is straightforward to show 
that (\ref{j+2}) can be written in a form analogous to (\ref{j-}) as
\begin{equation}
j_{\mu}^{(+)}= N_{(-)} \frac{i}{2} [\phi^{\dagger} \apm \phi +
\phi \apm \phi^{\dagger}].  
\end{equation} 

The results above can be summarized by defining the currents
\begin{equation}
q_{\mu}^{(\pm)}=\frac{i}{2} [ \phi^{\dagger} \apm \phi \pm 
\phi \apm \phi^{\dagger} ],
\end{equation}
which leads to
\begin{equation}
j_{\mu}^{(\pm)}= N_{(\mp)} q_{\mu}^{(\pm)}.
\end{equation}
The current $q_{\mu}^{(+)}$ vanishes, but the current $N_{(-)}q_{\mu}^{(+)}$ 
does not vanish.

The results above can be easily generalized to the case in which 
the field interacts with a background electromagnetic field, in a 
way similar to that in \cite{nikol_long}. The equations are essentially 
the same, but the derivatives $\partial_{\mu}$ are replaced by the 
corresponding gauge-covariant derivatives and the particle 2-point functions 
$W^{(P)\pm}$ are not equal to the antiparticle 2-point functions 
$W^{(A)\pm}$.

Similarly to the gravitational case, in the case of interaction with 
an electromagnetic background, different choices for the 2-point 
functions exist \cite{nikol_long}. 
One is a generalization of (\ref{W+-}) based 
on a particular choice of a complete orthonormal set of solutions to the 
equations of motion. The other is a generalization of (\ref{Wcreate}) and 
leads to a local description of the particle-antiparticle pair creation 
consistent with the conventional global description based on the 
Bogoliubov transformation. 
The third choice is based on the Schwinger-DeWitt Green function 
and leads to the conservation of the 
particle currents in classical electromagnetic backgrounds.  

The results of this section can also be generalized to 
anticommuting fermion fields (see also \cite{dolby}). 
As the analysis is very similar to the case of complex scalar fields, 
we simply note the final results. The particle and  
antiparticle currents 
can be written in a form similar to (\ref{curPA}) \cite{nikol_long}. 
In particular, a similar integration over $x'$ occurs, which is related 
to the extraction of $\psi^{(P)+}$, $\psi^{(P)-}$, 
$\psi^{(A)+}$, and $\psi^{(A)-}$ from the fermion fields 
$\psi$ and $\bar{\psi}$. Introducing the currents
\begin{equation}
q_{\mu}^{(\pm)}=\frac{1}{2}[\bar{\psi}\gamma_{\mu}\psi \pm 
\psi^T\gamma_{\mu}^T\bar{\psi}^T ],
\end{equation}
the currents
\begin{equation}
j_{\mu}^{(\pm)}=j_{\mu}^{(P)}\pm j_{\mu}^{(A)}
\end{equation}
can be written as
\begin{equation}
j_{\mu}^{(\pm)}= N_{(\pm)} q_{\mu}^{(\pm)}.
\end{equation}
The current $q_{\mu}^{(+)}$ vanishes
(due to the anticommutation relations among the fermion fields),
but the current $N_{(+)}q_{\mu}^{(+)}$ does not vanish. 
The current $j_{\mu}^{(-)}$ can also be written in more familiar forms as
\begin{eqnarray}
j_{\mu}^{(-)} & = & \, :\! \bar{\psi}\gamma_{\mu}\psi \! : \nonumber \\
 & = & \frac{1}{2}[\bar{\psi}\gamma_{\mu}\psi -
\psi^T\gamma_{\mu}^T\bar{\psi}^T ].
\end{eqnarray}

\section{Relation between the particle current and the 
energy-momentum tensor}\label{ENERGY} 
 
In classical field theory, the energy-momentum tensor of a real 
scalar field is 
\begin{equation}\label{Tmunu}
T_{\mu\nu}=(\partial_{\mu}\phi)(\partial_{\nu}\phi)-
g_{\mu\nu}\frac{1}{2}[g^{\alpha\beta}(\partial_{\alpha}\phi)
(\partial_{\beta}\phi)-m^2\phi^2] .
\end{equation}  
Contrary to the conventional concept of particles in quantum field theory, 
the energy-momentum is a local quantity. Therefore, the relation between 
the definition of particles and that of the energy-momentum is not clear 
in the conventional approach to quantum field theory in curved spacetime 
\cite{bd}. In this section, we exploit our
local and covariant description of particles to find a clearer relation 
between particles and their energy-momentum. 

In quantum field theory, one has to choose some ordering of the operators 
in (\ref{Tmunu}), just as a choice of ordering is needed in order to 
define the particle current. Although it is not obvious how to choose 
these orderings, it seems natural that the choice of ordering for one 
quantity determines the ordering of the other one. For example, if the 
quantum energy-momentum tensor is defined as 
$:\! T_{\mu\nu}\! :\, =N_{(+)} T_{\mu\nu}$, then the particle current 
should be defined as $N_{(-)} i\phi\apm\phi$. The nonlocalities 
related to the extraction of $\phi^+$ and $\phi^-$ from 
$\phi$, needed for the definition of the normal orderings 
$N_{(+)}$ and $N_{(-)}$, appear both in the energy-momentum and in the 
particle current.  
Similarly, if $W^{\pm}$ is chosen as in (\ref{W+-}) for one quantity, 
then it should be chosen in the same way for the other one. 
The choices as above lead to a consistent picture in which both the 
energy and the number of particles vanish in the vacuum $|0\rangle$ 
defined by $a_k|0\rangle =0$.

Alternatively, if $W^{\pm}$ is chosen as in (\ref{Wcreate}) for the 
definition of particles, then it should be chosen in the same way 
for the definition of the energy-momentum.
Owing to the extra time dependence, it is clear that both the
particle current and the energy-momentum tensor are not 
covariantly conserved in this case:
\begin{equation}\label{noncon12}
\nabla^{\mu}j_{\mu}\neq 0, \;\;\; 
\nabla^{\mu}T_{\mu\nu}\neq 0.
\end{equation}
While the first equation in (\ref{noncon12}) 
is exactly what one might want to obtain, the second one   
represents a problem. To be more specific,
assume, for simplicity, 
that spacetime is flat at some late time $t$.  
In this case, the 
normally ordered operator of the total number of particles 
at $t$ is 
\begin{equation}\label{creatN}
N(t)=\sum_{{\bf q}} A^{\dagger}_{{\bf q}}(t) A_{{\bf q}}(t), 
\end{equation}
(see (\ref{e10})), while the normally ordered operator of energy 
is 
\begin{equation}\label{creatH}
H(t)=\sum_{{\bf q}} \omega_{{\bf q}} A^{\dagger}_{{\bf q}}(t) 
A_{{\bf q}}(t).  
\end{equation} 
From (\ref{creatN}) and (\ref{creatH}) it is clear that 
the produced energy exactly corresponds to the produced particles.
A similar analysis can be done for the particle-antiparticle pair
creation caused by a classical electromagnetic background.  
Since the energy should be conserved, this suggests that 
$W^{\pm}$ should not be chosen as in (\ref{Wcreate}), i.e., that 
{\em classical backgrounds do not cause particle creation}.
Of course, in a time dependent gravitational field, the energy
of matter does not need to be conserved
in the ordinary, noncovariant sense; only the sum of matter
and gravitational energy should be conserved. However, in the
specific case above, the spacetime is flat at the late time $t$,
so the gravitational energy is zero. We can choose that the
metric at this late time is equal to the metric at the initial
time (at which the number of particles iz zero), such that
the time dependence of the metric at the intermediate
times is nontrivial. In such a case the contradiction between
particle creation and energy conservation is obvious. 

Of course, it is possible that the total 
energy-momentum is conserved owing to 
some mechanism of the back reaction that is not included in 
our calculation. 
However, just as the back reaction may prevent the creation of energy, 
it might also prevent the creation of particles. To support this idea, 
let us discuss a particular example. Consider a static electric field, 
the source of which is a stable charged particle. Various semiclassical 
calculations, based on the approximation that the electric field 
is static and 
classical, lead to pair creation. However, it is clear that  
pair creation is inconsistent with energy conservation. If a pair is 
really created, a back-reaction mechanism should provide the conservation 
of energy. One possibility is that the effect of the back reaction 
reduces to a modification of the electric field. However, 
the new electric field 
should be consistent with the Maxwell equations, so it is easy to see 
that it is impossible that the electric field is modified in a way 
consistent with energy conservation if the source 
of the field is not modified. On the other hand, the source cannot 
be modified as 
it is, by assumption, a stable particle. (The particle stability is a 
quantum property, so one cannot study it using semiclassical methods.) 
Therefore, we must conclude that, in this particular example,  
the back reaction completely prevents the pair creation. This demonstrates 
that a semiclassical treatment of the particle creation may lead to a  
completely wrong result. The formal results obtained in 
\cite{nikol_PLB,nikol_long} and this paper suggest 
a different semiclassical 
approximation according to which quantum particles are never created by 
classical backgrounds. In this approximation, the particles are defined 
by using a modified Schwinger-DeWitt Green function \cite{nikol_long}
to choose the 2-point functions 
$W^{\pm}$. Contrary to other semiclassical approximations, such a 
semiclassical approximation is self-consistent in the sense that 
there is no particle creation that violates the energy-momentum 
conservation law and particles are defined in a unique way without need 
to choose a particular time-coordinate and a particular gauge.  

Let us now choose the symmetric ordering and assume that spacetime is 
flat. As already discussed in Sec.~\ref{PARTICLE}, 
both the vacuum energy and the
vacuum number of particles are nonvanishing in this case. 
Taking the Lorentz-invariant normalization of the 
field in an infinite volume
\begin{equation}
\phi(x)=\int\frac{d^3k}{(2\pi)^32\omega({\bf k})} [a({\bf k})e^{-ikx}
+a^{\dagger}({\bf k})e^{ikx}],
\end{equation}
it is straightforward to show that the vacuum-expected value of the 
energy-momentum tensor is
\begin{equation}\label{tmunu_vac}
\langle 0|S_{(+)}T_{\mu\nu}|0\rangle =\frac{1}{2}\int\frac{d^3k}{(2\pi)^3}
\frac{k_{\mu}k_{\nu}}{\omega({\bf k})}.   
\end{equation}
Similarly, for the particle current we find 
\begin{equation}\label{jmu_vac}
\langle 0|S_{(-)}i\phi\apm\phi |0\rangle =\frac{1}{2}\int\frac{d^3k}{(2\pi)^3}
\frac{k_{\mu}}{\omega({\bf k})}.
\end{equation}
Note that the right-hand side of (\ref{jmu_vac})
is not only the expected value, but also the eigenvalue of
$S_{(-)}i\phi\apm\phi$ in the vacuum.
One can also define the energy-momentum current $T_{\mu}=n^{\nu}T_{\mu\nu}$, 
where $n^{\nu}$ is a unit timelike vector. We work in coordinates in 
which $n^{\nu}=(1,0,0,0)$, so in these coordinates
\begin{equation}\label{tmu_vac}                     
\langle 0|S_{(+)}T_{\mu}|0\rangle =\frac{1}{2}\int\frac{d^3k}{(2\pi)^3}
k_{\mu}.
\end{equation}
It is often argued that $\langle 0|S_{(+)}T_{\mu\nu}|0\rangle$ contributes 
to the cosmological constant. However, the presence of a cosmological 
constant is equivalent to an energy-momentum tensor of the form 
\begin{equation}\label{cosm}
T_{\mu\nu}^{{\rm cosm}}=\lambda g_{\mu\nu}.
\end{equation}
The right-hand side of (\ref{tmunu_vac}) does not have the 
form (\ref{cosm}). (For example, $T_{00}^{{\rm cosm}}$ and
$T_{11}^{{\rm cosm}}$ have the opposite sign, which is not the case
for (\ref{tmunu_vac})).
Actually, only the first term in (\ref{Tmunu}) contributes to
(\ref{tmunu_vac}), while the term proportional to $g_{\mu\nu}$
does not contribute to (\ref{tmunu_vac}). If the term
$m^2\phi^2$ in (\ref{Tmunu}) is replaced with a nontrivial
potential $V(\phi)$ which has a nonzero minimum at some
$\phi=\phi_{{\rm vac}}\neq 0$, then the term proportional to $g_{\mu\nu}$
contributes to (\ref{tmunu_vac}) even at the classical level,
which is the basic idea of quintessence models.
However, in the case we discuss 
$\phi_{{\rm vac}}=\langle 0|\phi|0\rangle =0$,
so there is no term proportional to $g_{\mu\nu}$.

Of course, the right-hand sides of (\ref{tmunu_vac})-(\ref{tmu_vac}) 
are infinite and should be regularized (and renormalized). 
However, there are many 
kinds of regularization and different kinds of regularizations are not always 
physically equivalent. One has to choose the regularization such that it 
preserves some physical property of unregularized expressions. 
For example, the cut-off regularization preserves the correspondence 
between the vacuum energy and the vacuum number of particles, but 
does not preserve the Lorentz invariance. On the other hand,
the dimensional regularisation and the zeta-function regularization 
preserve the Lorentz invariance, but do not preserve the correspondence 
between the vacuum energy and the vacuum number of particles. 
Therefore, in order to obtain regularized expressions, 
it seems necessary to abandon one of these two physical 
properties. The question is: Which one?   

It is widely believed that the vacuum energy-momentum should have the 
form (\ref{cosm}) because the vacuum should be relativistically invariant.
The Casimir effect \cite{iz} suggests that the vacuum energy 
should not be simply removed by the normal ordering. 
Therefore, the infinite vacuum energy-momentum is often renormalized 
such that it is required that the renormalized vacuum energy-momentum 
should have the form (\ref{cosm}). However, this requirement leads to a 
strange and counterintuitive result that the vacuum energy 
$(2\pi)^{-3}\int d^3k\,\omega({\bf k})/2$  
vanishes for massless fields and does not vanish for 
massive fields \cite{akhm}.

The discussion above suggests another possibility worthwhile to explore:
Perhaps, the requirement that the vacuum 
should be a relativistically invariant state with an energy-momentum 
of the form (\ref{cosm}) should be abandoned. Instead, the vacuum 
should be viewed in a way similar to the original Dirac's picture, 
in which the vacuum is filled not only with energy, but also 
with {\em particles} that carry this energy. 
The relativistic noninvariance of the vacuum reflects  
in flat spacetime as the existence of a preferred Lorentz 
frame in which the average velocity of vacuum particles is zero.
Indeed, if the 
space components $T_i$ of the energy-momentum vector vanish in 
a particular Lorentz frame but the time component $T_0$ does 
not vanish, then the space components $T_i$ do not 
vanish in another Lorentz frame. This is also consistent with 
the explicit expression for $T_i$  
in (\ref{tmu_vac}), because, if the limits of integration over 
space components of $k$ are symmetric in one Lorentz frame, then 
they are not symmetric in another Lorentz frame, so the 
contributions from the opposite $k_i$'s cancel only in one 
Lorentz frame. The same is true for the particle current 
(\ref{jmu_vac}) and for the nondiagonal components of 
(\ref{tmunu_vac}). 

The interpretation above of the vacuum energy requires the 
existence of a preferred frame. Although this may look strange 
from a theoretical point of view, it is an observed fact that 
a preferred frame exists in the Universe. This is the frame 
with respect to which the expanding Universe is homogeneous and 
isotropic at large space scales. According to the interpretation 
above, the vacuum energy (not related to a nonvanishing 
$\phi_{{\rm vac}}$) behaves as vacuum {\em matter}. 
However, this matter does not form structures (such as stars or 
galaxies), but is homogeneously distributed in the Universe. Such 
a nonclassical behavior can be understood as a consequence 
of the entanglement related to a very special quantum 
ground state $|0\rangle$. 
(This is similar to various nonclassical collective effects,
such as superconductivity,  
that appear in low-temperature solid-state physics.)

It is known that about 70\% of all energy 
in the Universe does not form structures. It is also known that 
the Universe expansion accelerates, which suggests that part of 
the energy has the cosmological-constant form (\ref{cosm}) with a 
negative pressure. (For a review, see, e.g., \cite{tur,bah}.) 
Future more precise measurements of the negative pressure that 
causes acceleration and of the energy-density that does not 
form structures might show that not all energy that does not 
form structures can be explained by a cosmological constant, 
which would be an (indirect) experimental 
confirmation that the picture of 
the vacuum proposed above is qualitatively correct. 
To obtain a respectable quantitative picture, the  
renormalization is necessary. In particular, the 
running of the vacuum energy \cite{sola,guber} might provide 
useful information that can be compared with experiments.
 
\section{Conclusion}\label{CONCL}

In this paper, it is shown that the recently discovered 
particle currents \cite{nikol_PLB,nikol_long} can be written in  
purely local forms. The nonlocalities are hidden in the extraction 
of $\phi^+$ and $\phi^-$ from $\phi$. The formalism is applied 
to a local description of particle creation by gravitational
backgrounds, with emphasis on the description
of particle creation by black holes. The formalism also reveals 
a relation between particles and their energy-momentum, which 
suggests that it might not be consistent to use semiclassical methods
for a description of particle creation. The relation between particles and
their energy-momentum also suggests that the vacuum energy might contribute 
to dark matter that does not form structures, instead of contributing 
to the cosmological constant. 

\section*{Acknowledgement}
The author is grateful to H.~\v Stefan\v ci\'c for useful comments and
suggestions and to B.~Guberina for a comment on the running of the 
vacuum energy.
This work was supported by the Ministry of Science and Technology of the
Republic of Croatia under Contract No.~0098002.

\end{document}